\documentclass[twocolumn,english,showpacs,aps,pra,floats]{revtex4}
\usepackage[]{fontenc}
\usepackage[latin1]{inputenc}
\usepackage{graphicx,subfigure,dcolumn,bm}
\usepackage{amsfonts,amsmath,amssymb,epsfig,babel}
\newcommand{\bra}[1]{\langle #1 |}
\newcommand{\ket}[1]{| #1 \rangle}
\newcommand{\be}{\begin{equation}}
\newcommand{\ee}{\end{equation}}
\newcommand{\beq}{\begin{eqnarray}}
\newcommand{\eeq}{\end{eqnarray}}
\newcommand{\bi}{{\bf i}}
\newcommand{\bj}{{\bf j}}

\makeatother

\begin{document}

\title{Detecting the tunneling rates for strongly interacting fermions on optical lattices}
\author{Alberto Anfossi$^{1,2}$, Luca Barbiero$^1$, and Arianna Montorsi$^1$}
\affiliation{$^1$ Dipartimento di Fisica del Politecnico, corso Duca degli Abruzzi 24, I-10129, Torino, Italy}
\affiliation{$^2$ Dipartimento di Fisica dell'Universit\`a di Bologna, viale Berti-Pichat 6/2, I-40127, Bologna, Italy}

\date{\today}

\begin{abstract}
Strongly interacting fermionic atoms on optical lattices are studied through a Hubbard-like model Hamiltonian, in which tunneling rates of atoms and molecules between neighboring sites are assumed to be different. In the limit of large on-site repulsion $U$, the model is shown to reproduce the $t$-$J$ Hamiltonian, in which the $J$ coefficient of the Heisenberg term depends on the particle-assisted tunneling rate $g$: explicitly, $J=4 g^2/U$. At half-filling, $g$ drives a crossover from a Brinkman-Rice paramagnetic insulator of fully localized atoms ($g=0$) to the antiferromagnetic Mott insulator of the standard Hubbard case ($g=t$). This is observed already in the number of doubly occupied sites under the intermediate coupling regime, thus providing a criterion for extracting from measurements the effective value of $g$.
\end{abstract}

\maketitle

\section{Introduction}

In recent years, the field of ultracold atoms has proven to be a very rich and growing research area \cite{GPS}, one of the basic motivations residing in the possibility of experimental access to and probing of a wide range of prominent problems in condensed-matter physics. For example, the rapid development of experiments on ultracold atomic gases loaded onto optical lattices has supported the observation of the superfluid-Mott-insulator transition in the Bose-Hubbard model \cite{GMEHB}.

One of the most relevant challenges in this field turns out to be the simulation of the fermionic Hubbard model \cite{JAZO,LEal} and its extensions \cite{Camp,Hirsch}, which are  believed to be the major candidates in describing high-temperature superconductivity. Most recently, in very intriguing experiments with fermionic  $^6$Li \cite{Jetal} and $^{40}$K \cite{Bloch} atoms loaded onto optical lattices, a Mott-insulating phase was found under the regime of strong on-site repulsion between particles, characterized by the presence of one atom per site and by a number of doubly occupied sites (doublons), depending on the strength of the external parameters. Such a state turns out to be relatively easy to achieve experimentally --due to its incompressibility-- and its observed behavior is qualitatively consistent with well-established properties of the Hubbard Hamiltonian \cite{Hub}.  In addition, it has been pointed out that under a regime of strong interaction, the necessary Feshbach resonance could induce highly nontrivial processes, which involve multiband populations and off-site interactions. These are  described by a generalized Hubbard Hamiltonian proposed by Duan \cite{Duan}, previously known in the literature as the Simon-Aligia model \cite{SiAl}. In fact, this model differs form the standard Hubbard model in the effective tunneling rates between neighboring sites, which are assumed to be dependent on the type of particles involved, at variance with the standard case. The fact that in experiments it is possible to control separately the dynamics of molecules (doublons) and that of single atoms is explicitly included in the Hamiltonian. Here we show that this fact can be used to induce a strong coupling regime already at moderate repulsive interaction $U$. Such a regime can possibly be observed in experiments, for instance, as a lower number of doublons in the insulating phase, with respect to the standard Hubbard case.

In the following, we first investigate analytically the strong-coupling regime for the Duan Hamiltonian at arbitrary tunneling rates and fillings. Then we focus on the half-filled case and explore --also numerically by means of DMRG simulations-- the behavior of the insulating state. The ultimate aim is to infer from experimental data the effective value of the tunneling rates in real systems.

\section{Lattice Model Hamiltonian}

\begin{figure}
    \includegraphics[width=70mm,keepaspectratio,clip]{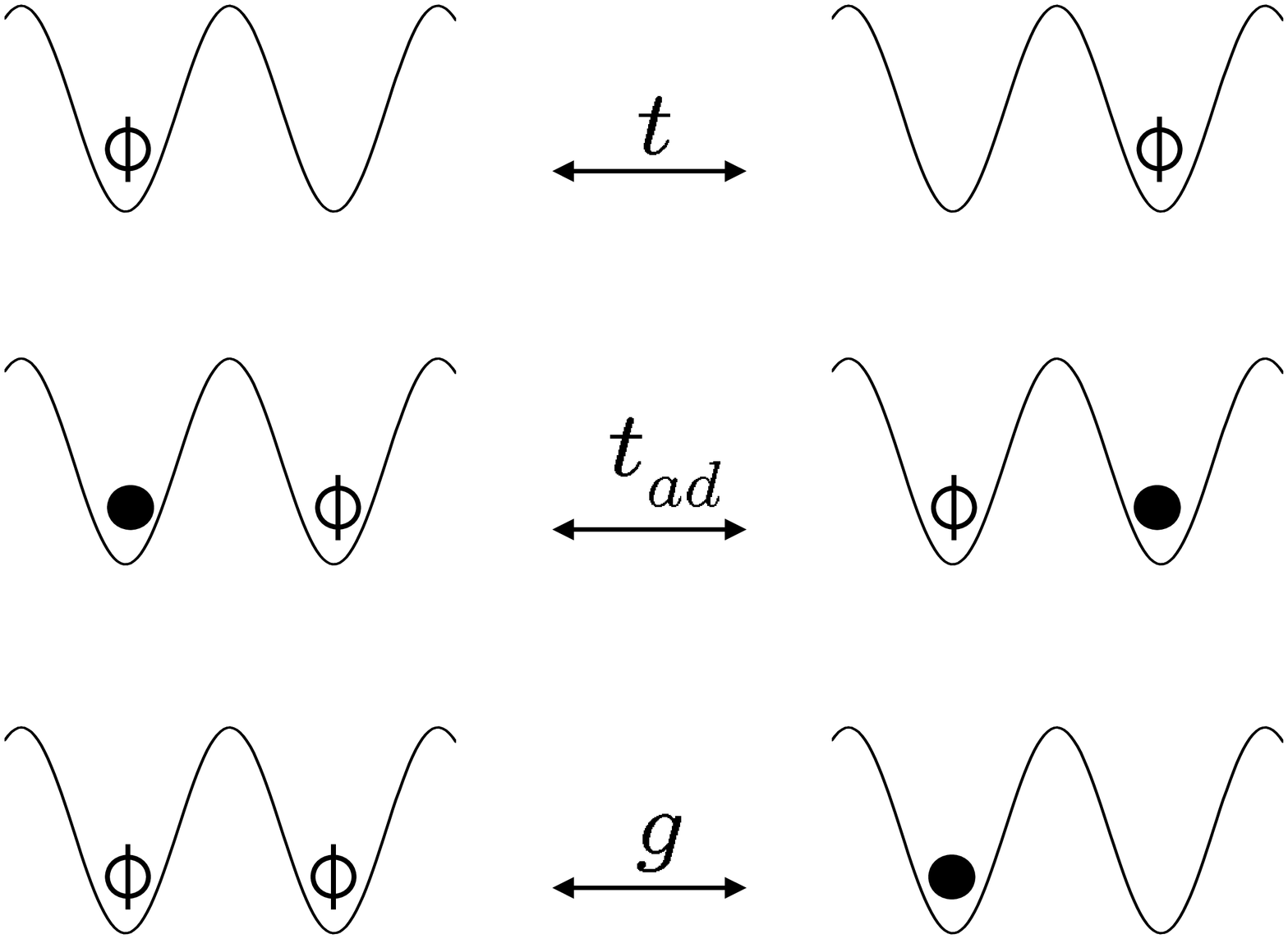}
    \caption{Processes between neighboring sites described by the three independent tunneling rates $t$, $g$, and $t_{ad}$. Here slashed and filled circles represent single atoms (of arbitrary spin orientation) and for diatomic molecules (doublons), respectively.}\label{fig1}
\end{figure}

The model Hamiltonian reads
\begin{widetext}
\be
    H =  -\sum_{\langle \bi \bj \rangle,\sigma} \left [t+\delta g(n_{\bi\bar{\sigma}} +n_{\bj\bar{\sigma}})+\delta t n_{\bi\bar{\sigma}}n_{\bj\bar{\sigma}}\right ] c_{\bi\sigma}^{\dagger}c_{\bj\sigma} + \frac{U}{2}\sum_{\bi} n_{\bi}( n_{\bi}-1)\;,\label{ham_SA}
\ee
\end{widetext}
where \textbf{$\sigma=\pm 1$ ($\bar{\sigma}\doteq -\sigma$)} identifies the two internal states of the fermionic atoms, $\langle ij \rangle$ denotes two neighboring sites on a $d$-dimensional regular lattice and $n_i\doteq n_{i+}+ n_{i-}$. Here $t$  describes the direct hopping of atoms of a given population between neighboring  sites, while $\delta g=g-t$ and $\delta t=t+t_{ad}-2 g$ are the deviation from the direct hopping case (in which $\delta g=0=\delta t)$, induced by correlations in proximity of a wide Feshbach resonance. More precisely, as shown schematically in Fig. \ref{fig1}, $g$ describes the tunneling configuration in which one atom is transferred to a site already occupied by an atom of a different species; $t_{ad}$ accounts for the motion of one atom between two already-occupied sites, thus exchanging a molecule and a fermionic atom located at neighboring sites. Finally, $U$ is the energy cost of the molecule, which works as an effective detuning parameter.

The derivation of Hamiltonian (\ref{ham_SA}) for a system of ultracold fermionic atoms trapped on an optical lattice has been given in \cite{Duan} and \cite{DuZo}. In the case of a double-well lattice, the numerical solution for two fermions interacting across a Feshbach resonance was also used to infer the range of parameters in which $H$ works as an effective Hamiltonian \cite{KeDu}.

In one dimension, the ground state of $H$ has been investigated in detail in the context of high $T_c$ superconducting materials  for the choice $\delta t=0$ (see, e.g., Ref. \cite{AAA} and references therein). Some results have also been achieved at $\delta t\neq 0$ \cite{NAKA, MON,ADM}. Most recently the case of imbalanced atoms has also been explored \cite{ABM,Duan}. The ground-state phase diagram in the moderate interaction regime $U\leq 4 t$ differs substantially from what one would get by considering the standard Hubbard model. In particular, for $\delta g>t/2$ and $U<U_c (g)$, it is characterized by a phase that is superconducting even at half-filling and that displays an enhanced number of doublons per site, $n_d$. By increasing the population imbalance, $n_d$ remains unchanged up to a critical polarization $p_c$, beyond which some pair breaks and macroscopic phase segregation takes place \cite{ABM}. At half-filling, the presence of such a superconducting phase shifts the insulating behavior characteristic of the Hubbard model to higher $U$ values, $U>U_c(g)>0$.

In the following, we investigate how the actual $\delta g$ value also affects the behavior of the system in the intermediate to strong coupling region $U>U_c$. In fact, it will turn out that the choice of $\delta t$ is irrelevant to the strong coupling behavior.

\section{The strong-coupling limit}

To perform the strong-coupling limit of Eq. (\ref{ham_SA}), we generalize to the present case a procedure developed in Ref. \cite{MGY} for the Hubbard model. As the first step, we rewrite Hamiltonian (\ref{ham_SA}) in terms of Hubbard projectors $X^{\alpha\beta}_\bj=\ket{\alpha}_{\bj \bj}\bra{\beta}$, where $\ket{\alpha}_\bj$ are the four states spanning the basis of the vector space at a given site ${\bf j}$, with  $\alpha=\{0,+,-,2\}$ and $\ket{2}_\bj\doteq \ket{+-}_\bj$. Explicitly,
\begin{eqnarray}
    H &=& U\sum_\bi X_\bi^{22}-\sum_{<\bi\bj>\sigma}\Big[t X^{\sigma
    0}_\bi X^{0\sigma}_\bj+t_{ad} X^{2\sigma }_\bi X^{\sigma2}_\bj
    \nonumber\\
    &&+g (-)^\frac{1+\sigma}{2} \left(X^{2\sigma }_\bi X^{0\bar{\sigma}}_\bj+X^{\bar\sigma
    0}_\bi X^{\sigma 2}_\bj\right)\Big] \label{HX} \; .
\end{eqnarray}
We observe that the hopping terms in Eq. (\ref{HX}) can be classified according to the change in the number of fermion pairs involved:
\begin{eqnarray}
    T_0 &=& \sum_{<\bi\bj>\sigma}\left (t X^{\sigma 0}_\bi X^{0\sigma}_{\bj}+t_{ad} X^{2\sigma }_\bi X^{\sigma2}_{\bj}\right ) =T_0^{(0)}+t_{ad} T_0^{(2)}\;,\nonumber\\
    T_{+1} &=& \sum_{<\bi\bj>\sigma}(-1)^\frac{1+\sigma}{2} X^{2\sigma }_\bi X^{0\bar{\sigma}}_{\bj}\;,\nonumber\\
    T_{-1} &=& T_{+1}^\dagger\;. \label{T_nd}
\end{eqnarray}
More precisely, $T_0$ leaves unchanged the number of local pairs whereas $T_{+1(-1)}$ creates (destroys) a pair. It is important to note that $T_m^\dagger=T_{-m}$ and $\left[N_d,T_m\right]=mT_m$, where $N_d= \sum_\bi  X_\bi^{22}$, its expectation value being the total number of doublons in the system. In terms of the above operators, the Hamiltonian (\ref{HX}) reads: $H=U N_d +T_0+g (T_{+1}+T_{-1})$.

In the strongly interacting regime $t/U<<1$ we may decompose the fermionic Hilbert space of the model into subspaces characterized by different values of $N_d$:
\begin{equation}
    \mathcal{H}=\bigoplus_{N_d=0}^{N/2}\mathcal{H}_{N_d} \; .
\end{equation}

Indeed, in this case the energy spectrum of $H$ splits into well-separated subbands with different $N_d$, and a gap $(\sim U)$ opens between them. Hence, within this limit, we consider an effective Hamiltonian which does not mix different sectors of the Hilbert space, namely, conserving the total number of fermionic pairs. To obtain such a Hamiltonian we perform a unitary transformation that eliminates to the lowest order in $U$ (the zeroth order) the terms not commuting with $N_d$. Explicitly,
\begin{equation}
    S=S^{(1)}=\frac{g}{U }(T_{+1}-T_{-1}) \; ,
\end{equation}
so that, to the first order in $U^{-1}$, the rotated Hamiltonian $H^{(1)}$ reads
\begin{eqnarray*}
    H^{(1)}&=& e^S He^{-S} \nonumber \\
    &=& H+ \frac{[S,H]}{1!}+\frac{\left[S,[S,H]\right]}{2!}+... \nonumber\\
    &=& U N_d+T_0 + U^{-1}\left(g^2 [T_{+1},T_{-1}]\right.\nonumber \\
    &&\left. +g [T_{0},T_{-1}]+ g [T_{+1},T_{0}]\right) \; .
\end{eqnarray*}
Now one can think to iterate this procedure, eliminating the terms not conserving $N_d$ also to the first order in $U^{-1}$. This is implemented by choosing
\begin{equation}
    S^{(2)}=g U^{-2}\left([T_{+1},T_{0}] - \mbox{h.c.}\right)\;,
\end{equation}
so that the rotated Hamiltonian, up to second order in $U^{-1}$, reads
\begin{equation}
    H^{(2)}=U N_d +T_0 -\frac {2 g^2}{U} [T_{-1},T_{+1}] + \mathcal{O}(U^{-2})\;. \label{hamsc}
\end{equation}
The effective Hamiltonian in the lowest-energy sector ({\it i.e.}, $N_d=0$) is hence obtained as
\begin{equation}
    H_{eff}=T^{(0)}_0-J\sum_{<ij>}\left(\bar{S_i}\bar{S_j}- \frac{1}{4}n_in_j\right) + O(U^{-2})\; , \label{hamtJ}
\end{equation}
where --as usual-- the three-site term has been neglected\cite{AUE}. Remarkably, $H_{eff}$ is still in the form of the $t$-$J$ model, as in the strong-coupling limit of the standard Hubbard case ($g=t$). However, in the present case the $J$ coefficient has changed to $J=4  g^2/U$. Note that by specializing to half-filling one recovers the antiferromagnetic Heisenberg model.

\section{Discussion}

Equation (\ref{hamtJ}) provides --in the strong-coupling limit-- the ground-state energy of the model Hamiltonian (\ref{ham_SA}) in arbitrary dimension, at arbitrary $g$ and $t_{ad}$ starting from the known results for the $t$-$J$ (generic filling) and Heisenberg (half-filling) models. Given the dependence of $J$ on $g$, Eq. (\ref{hamtJ}) in fact describes the crossover from an antiferromagnetic insulator to a Brinkman-Rice \cite{BRRI} paramagnetic insulator of fully localized fermions (at $g=0$). The effect was already predicted \cite{ADM} in one dimension by means of numerical investigation of the model of Eq. (\ref{ham_SA}), and here it finds its analytic confirmation.

Such a result can be viewed as a consequence of the interplay of the exchange term (with coefficient $\propto g^2/U$) and the kinetic term describing the motion of single atoms in a background of empty sites. The latter, which is also known as the infinite-$U$ Hubbard model, in one dimension describes the physics of a system of spinless fermions, with a vanishing number of doubly occupied sites, and energy per site $e=-2/\pi \sin{\pi n}$. In particular, at $n=1$ it describes an insulator in which each fermion is fully localized at a different lattice site --no matter its spin orientation: the Brinkman-Rice insulator. Increasing $g$ (and hence $J$) amounts to a gain energy, thanks to magnetic interactions, without affecting $e$: the insulator progressively increases its magnetic ordering and diminishes its energy, to reach the Hubbard limit value. Even for $N_d\neq 0$, inspection of Eq. (\ref{hamsc}) still shows that higher $g$ values correspond to lower energies, so that one expects the spinless character of the insulator to emerge (and $N_d$ to diminish) with decreasing $g$.

\begin{figure}
    \includegraphics[width=90mm,keepaspectratio,clip]{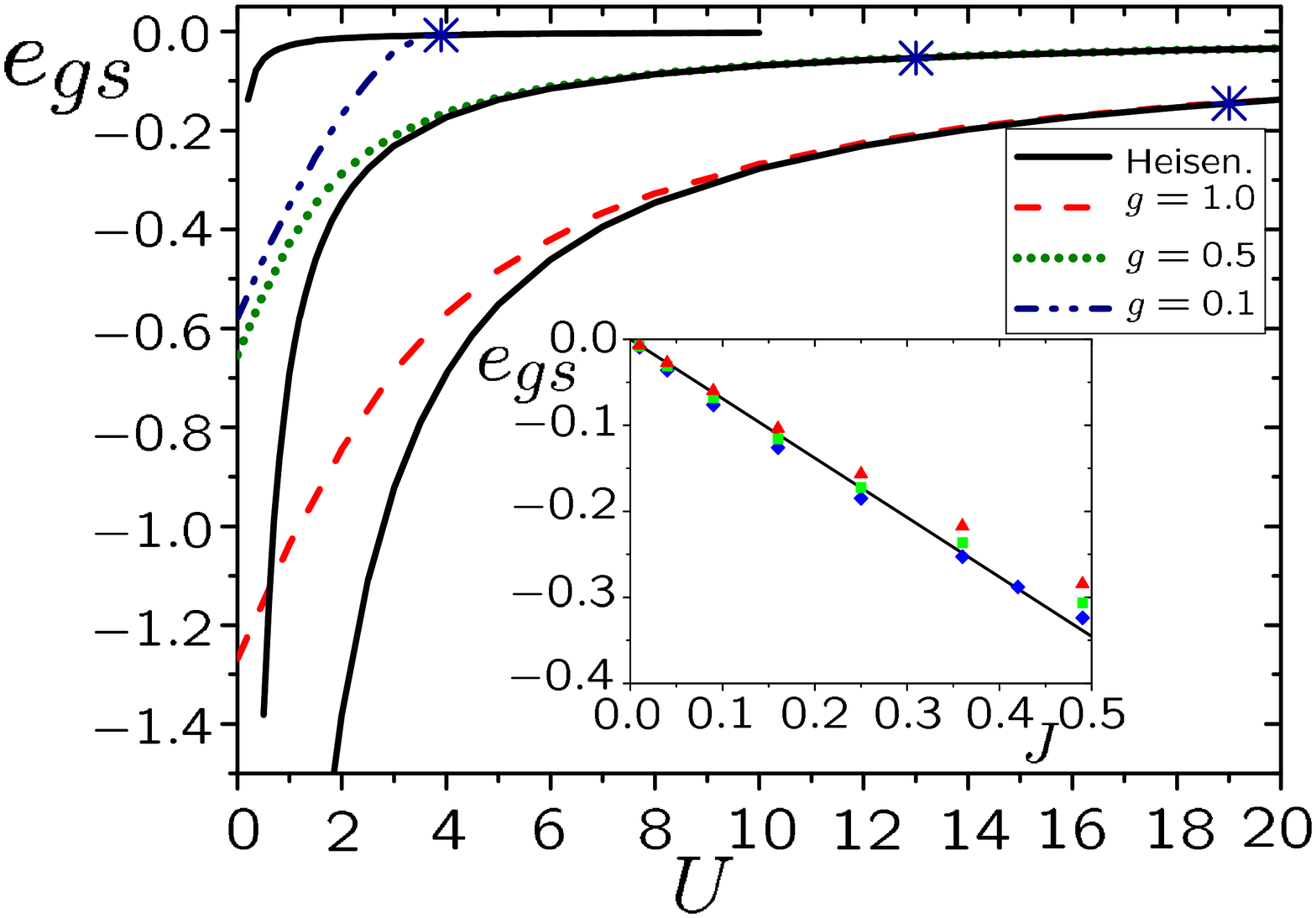}
    \caption{(Color online) Ground-state energy density $e_{gs}$ at half-filling, $t=1$, versus $U$ at three different $g$ values, as obtained by numerical simulations with DMRG on $L=80\div 120$ sites. Dotted, dashed, and dot-dashed lines represent the results obtained for Hamiltonian (1), whereas solid lines are the results obtained for the Heisenberg model at the same parameter values. Asterisks denote in each case the value of $U$ ($U_*$) at which the two energies differ by less than 1\%. Inset: $e_{gs}$ versus $J$ at $t_{ad}=0.8$ (blue diamonds), $t_{ad}=0.4$ (green squares), and $t_{ad}=-0.8$ (red triangles).}\label{fig2}
\end{figure}

Some implications of the strong-coupling limit, (\ref{hamsc}) and (\ref{hamtJ}), on the behavior of the system described by Eq. (\ref{ham_SA}) at intermediate values of the repulsive interaction $U$ are given here, with the help of numerical analysis. Since $J$ also depends on $g$, one may expect that, with decreasing $g$, the strong-coupling regime --identified as the regime characterized by Hamiltonian (\ref{hamtJ})-- is reached at lower values of $U$. This is shown in Fig. \ref{fig2}, where we report the ground-state energy for Hamiltonian (\ref{ham_SA}), compared to that of the corresponding $t$-$J$ limit, Eq. (\ref{hamtJ}), at different $g$ values and $t=1$. If we define $U_*$ as the value of the interaction at which the difference between the two energies is lower than 1\%, it is seen that in fact $U_*$ decreases with decreasing $g$.
This is in agreement with exact results on the two limiting cases known in the literature: while in the $g=0$ limit \cite{MON} the Brinkman-Rice insulator is the exact ground state at large enough finite $U>U_c$ (in one dimension, $U_c= 4 t$), for $g=t$ the Heisenberg antiferromagnet limit is reached exactly only at $U=\infty$. From an experimental point of view, this observation suggests that the strong coupling regime could be explored even at a not too large detuning $U$, in the case where $g$ is low enough.

Moreover, since $J$ in Eq. (\ref{hamtJ}) does not depend on $t_{ad}$ the same should happen for the ground-state energy versus $J$, at large enough $U$ values. This is shown in the inset in Fig. \ref{fig2}, where in one dimension the ground-state energy is plotted and compared to that of the Heisenberg model at different $t_{ad}$ values and fixed $U$. It is seen that already at a moderate value of the repulsive interaction $U$, the effect of $t_{ad}$ is in fact negligible.

\begin{figure}
    \includegraphics[width=70mm,keepaspectratio,clip]{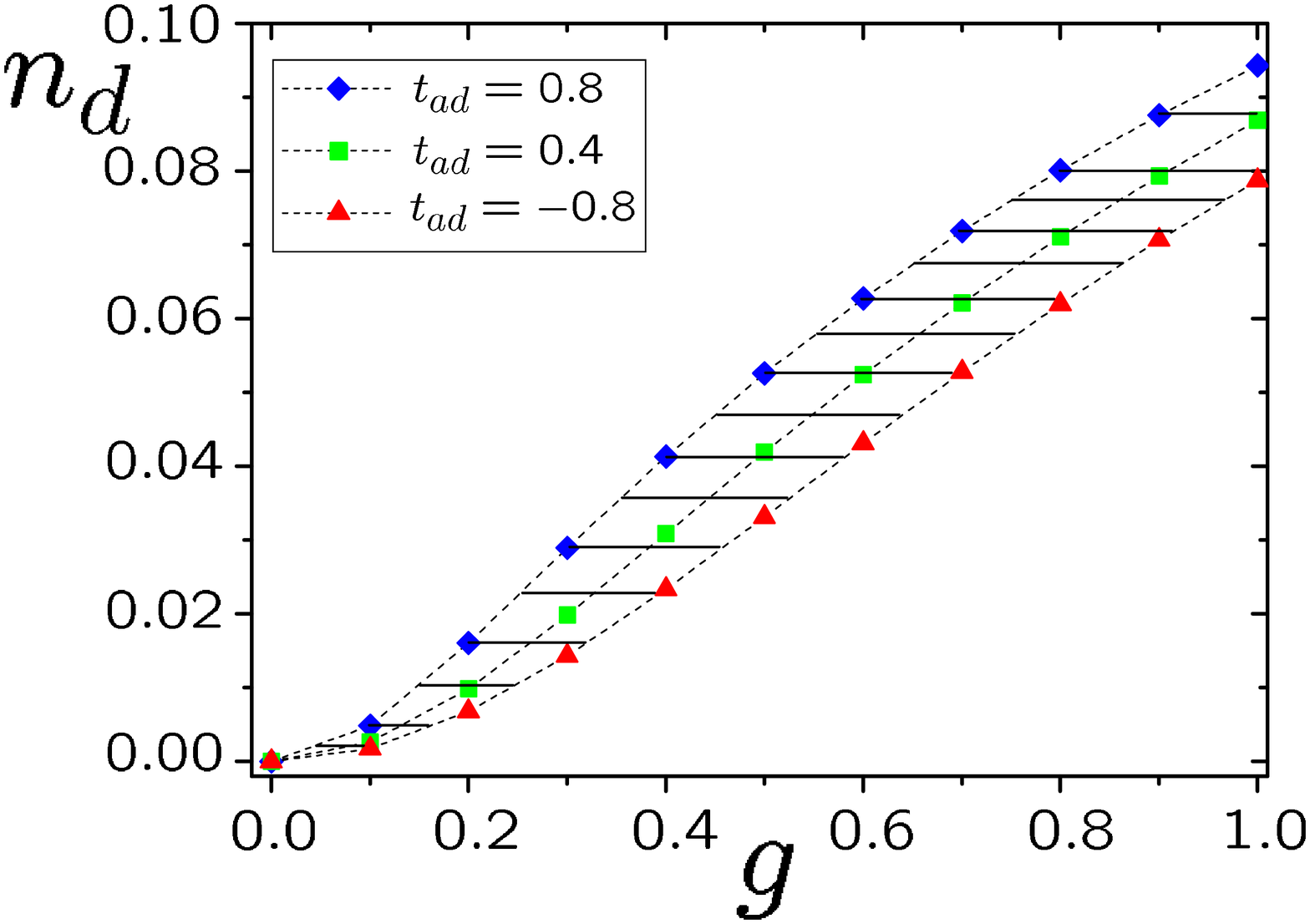}
    \caption{(Color online) DMRG results for the number of doubly occupied sites (doublons) as a function of $g$ at different $t_{ad}$ values, with $U=4t$. The solid horizontal lines represent the range of values of $g$ consistent with a given value of $n_d$. Dashed curves are guides for the eyes.}\label{fig3}
\end{figure}

In Fig. \ref{fig3} the role of $g$ for intermediate values of the interaction $U$ is exploited to investigate the dependence of the number of doublons $n_d=N_d/L$ on $U$. This analysis confirms that the strong coupling behavior also characterizes this regime, since $n_d$ still decreases by decreasing $g$, approaching smoothly the strong coupling value $0$ for $g=0$.  This result is consistent with the intuitive picture which emerges from the formulation given in (\ref{HX}) of the Hamiltonian $H$: the lower is $g$, the more processes requiring the presence of no doubly occupied sites (those with coefficient $t$) are favored with respect to those requiring $n_d\neq 0$ (with coefficients $g$ and $t_{ad}$). The result is, instead, in contrast with what happens in the moderate coupling region $U<U_c(g)$ \cite{ADM} where, as mentioned, $n_d$ could even increase slightly with decreasing $g$.

Since $n_d$ is typically measured in experiments, Fig. \ref{fig3} offers a straightforward way to extract information about the actual value of $g$ in the experimental setup. We highlight with horizontal straight lines the region in which the range of values of $g$ is compatible with a measured value of $n_d$ at a given $U$. Indeed, the range $|t_{ad}|>t$ can be mapped into that with $|t_{ad}|<t$ shown in figure. It is seen that at fixed $U$ the standard Hubbard case reaches the highest relative values of $n_d$.

Besides, lower values of $n_d$ (and hence of $g$) are consistent with experimental data in \cite{Bloch} at moderate values of the energy trap, assuming that the dimension of the lattice enters only through the bandwidth. Note that already in the case of a double well\cite{KeDu}, the value of $g$ consistent with the observed spectrum was found to be lower than 1 ($g\approx 0.8 t$). Moreover, in the case of real condensed matter systems, $g\approx 0.5 t$ \cite{Hub}.

\section{Conclusions}

In summary, we have shown that the strong-coupling regime of Hamiltonian (\ref{ham_SA}) is described by the $t$-$J$ Hamiltonian in which $J=4 g^2/U$. At half-filling, the effect of the particle-dependent tunneling rate $g$ is to drive the Mott insulator from an antiferromagnet toward a paramagnetic configuration of fully localized atoms. At intermediate values of the interaction $U$, this behavior induces a related behavior of the number of doubly occupied sites $n_d$, which decreases down to 0 with $g$, thus providing a criterion for inferring the effective value of $g$ directly from experimental measurements of $n_d$.

\begin{acknowledgments}
We thank F. Ortolani for the DMRG code. DMRG simulations of the Heisenberg chain were performed using the ALPS project libraries and applications \cite{ALPS}. This work was partly supported by Grant PRIN 2007JHLPEZ from the Italian MIUR.
\end{acknowledgments}

\end{document}